\documentclass[preprint,showpacs,showkeys,preprintnumbers,amsmath,amssymb]{revtex4}
\usepackage{graphicx}% Include figure files
\usepackage{dcolumn}% Align table columns on decimal point
\usepackage{bm}% bold math

\begin{document}

\title{Interference between the halves of a double-well trap containing a Bose-Einstein condensate
}
\author{Zhao-xian Yu}
\email{zxyu@eyou.com} \affiliation{ Department of Applied Physics,
University of Petroleum (East China), Dongying 257061, Shandong
Province, P.R.China}
\author{Zhi-yong Jiao}
\affiliation{1.Department of Applied Physics, University of
Petroleum (East China), Dongying 257061, Shandong Province,
P.R.China \\2. Photonics Center, College of Physics, Nankai
University, Tianjin 300071, P.R.China}

\begin{abstract}
Interference between the halves of a double-well trap containing a
Bose-Einstein condensate is studied. It is found that when the
atoms in the two wells are initially in the coherent state, the
intensity exhibits collapses and revivals, but it does not for the
initial Fock states.  Whether the initial states are in the
coherent states or in a Fock states, the fidelity time has nothing
to do with collision. We point out that  interference and its
fidelity can be adjusted experimentally by properly preparing the
number and initial states of the system.
\end{abstract}

\pacs{03.75.Fi, 05.30.Jp, 42.25.Hz} \keywords{ Interference,
fidelity, Bose-Einstein condensate} \maketitle

\section{\label{sec:intro}Introduction}

Since the recent experimental realization of Bose-Einstein
condensation in small atomic samples \cite{mh,cc,kb,mo}, there has
been much theoretical interest focused on the physical properties
and nature of Bose-Einstein condensed systems such as coherent
tunneling, the collapses and revivals in both the macroscopic wave
function and the interference patterns
\cite{yj,gj,em,tm,ab,ji,xx,wm}. It is hoped that the study of
those experimental systems will give new insight into the physics
of BEC. Since the current understanding of BEC is largely
influenced by the concept of a macroscopic wave function, the
study of this feature is of foremost importance. The investigation
of interference phenomena should be perfectly suited for this
purpose. Another motivation for the study of interference
properties is the envisioned development of a new source of atoms,
based on BEC, with high flux and coherence. It is expected to
stimulate atomic interference experiments.

Recently,  Javanainen et al. \cite{jj} have theoretically studied
atom number fluctuations between the halves of a double-well trap
containing a Bose-Einstein condensate, in which the two-mode
approximation is used, which assumes that only two one-particle
states are involved. They have developed an analytical
harmonic-oscillator-like model and verified numerically for both
stationary fluctuations in the ground state of the system, and for
the fluctuations resulting from splitting of a single trap by
dynamically erecting a barrier in the middle.

This paper is organized as follows. Sec.\ref{sec:mo} gives the
solution of model. Sec. \ref{sec:co} studies collapses and
revivals of interference intensity. Sec. \ref{sec:fi} investigates
fidelity of interference. A conclusion is given in the last
section.

\section{ \label{sec:mo}Model}

In a symmetric double-well potential, the ground state of a single
particle is represented by an even wave function $\psi_g$ that
belongs equally to both wells of the potential. Provided that
barrier between the halves of the potential is tall enough so that
the tunneling rate between the potential wells is small, nearby
lies an excited odd state $\psi_e$ that likewise belongs to both
halves of the double well. The starting point is that we take only
two one-particle states $\psi_g$ and $\psi_e$ to be available to
the $N$ bosons. The reason  for this choice of including only the
two lowest lying modes for the double-well potential is that the
other modes are energetically inaccessible.

We adopt the usual two-particle contact interaction
$U(\mathbf{r}_1,\mathbf{r}_2)=(4\pi\hbar^2a/m)\delta(\mathbf{r}_1-\mathbf{r}_2)$,
where $a$ is the s-wave scattering length and $m$ the atomic mass.
Given the restricted state space of precisely two one-particle
states, the many-particle Hamiltonian is [13]
$$H=\frac{1}{2}(\epsilon_1+\epsilon_2)(a^+_1a_1+a^+_2a_2)+\frac{1}{2}
(\epsilon_1-\epsilon_2)(a^+_1a_1-a^+_2a_2)$$
\begin{equation}
+K_{11}a^{+2}_1a^2_1+K_{22}a^{+2}_2a^2_2+K_{12}(a^{+2}_1a^2_2+a^{+2}_2a^2_1+4a^+_1a_1a^+_2a_2),
\end{equation}
where we set $\hbar=1$, and correspondingly use the terms energy
and (angular) frequency interchangeably. In Eq. (1) $a_1$ and
$a_2$ are the boson operators for the excited and ground wave
functions. The constants $\epsilon$ and $K$ are the one- and
two-particle matrix elements
\begin{equation}
\epsilon_1=\int
d^3\mathbf{r}\psi_e(\mathbf{r})[-\frac{1}{2m}\nabla^2+V(\mathbf{r})]\psi_e(\mathbf{r}),
\end{equation}
\begin{equation}
\epsilon_2=\int
d^3\mathbf{r}\psi_g(\mathbf{r})[-\frac{1}{2m}\nabla^2+V(\mathbf{r})]\psi_g(\mathbf{r}),
\end{equation}
\begin{equation}
K_{11}=\frac{2\pi a}{m}\int d^3\mathbf{r}|\psi_e(\mathbf{r})|^4,
\end{equation}
\begin{equation}
K_{22}=\frac{2\pi a}{m}\int d^3\mathbf{r}|\psi_g(\mathbf{r})|^4,
\end{equation}
\begin{equation}
K_{12}=K_{21}=\frac{2\pi a}{m}\int
d^3\mathbf{r}|\psi_e(\mathbf{r})|^2|\psi_g(\mathbf{r})|^2,
\end{equation}
$V(\mathbf{r})$ is the symmetric double-well binding potential.
Without restricting the generality, we assume that the wave
functions $\psi_{e,g}$ are real. To simplify the discussion, we
set $K_{12}=K_{11}=K_{22}=K$, and
$\epsilon_1=\epsilon_2=\epsilon$.

In order to solve Eq. (1), we introduce the following
transformations
\begin{equation}
a_1=\frac{1}{\sqrt{2}}(A_1e^{ikt}-iA_2e^{-iKt}),
\end{equation}
\begin{equation}
a_2=\frac{1}{\sqrt{2}}(A_1e^{ikt}+iA_2e^{-iKt}),
\end{equation}
where $[A_i,A^+_j]=\delta_{ij}$. We have from Eq. (1)
$$H=\epsilon(A_1^+A_1+A^+_2A_2)+K[(A_1^+A_1+A^+_2A_2)^2$$
\begin{equation}
-3(A_1^+A_1+A^+_2A_2)-3A_1^+A_1A^+_2A_2+(A_1^+A_1)^2+(A^+_2A_2)^2],
\end{equation}
If we define two bases as follows
\begin{equation}
|n,m)=\frac{1}{\sqrt{n!m!}}A_1^{+n}A_2^{+m}|0,0),
\end{equation}
\begin{equation}
|n,m>=\frac{1}{\sqrt{n!m!}}a_1^{+n}a_2^{+m}|0,0>,
\end{equation}
We have
\begin{equation}
H|n,m)=E_{n,m}|n,m),
\end{equation}
with
\begin{equation}
E_{n,m}=\epsilon(n+m)+K(2m^2+2n^2-mn-3n-3m).
\end{equation}

We now define two-mode coherent states as follows
\begin{equation}
|\alpha_1,\alpha_2>=D_{a_1}(\alpha_1)D_{a_2}(\alpha_2)|0,0>,
\end{equation}
\begin{equation}
|u_1,u_2>=D_{A_1}(u_1)D_{A_2}(u_2)|0,0>,
\end{equation}
where the displacement operators are defined by
\begin{equation}
D_{a_i}(\alpha_i)=\exp[\alpha_i^*a_i-\alpha_ia_i^+]~~(i=1,2),
\end{equation}
\begin{equation}
D_{A_i}(u_i)=\exp[u_i^*A_i-u_iA_i^+]~~(i=1,2).
\end{equation}
It is easy to see
\begin{equation}
|\alpha_1,\alpha_2>=|\frac{1}{\sqrt{2}}(\alpha_1+\alpha_2)e^{-iKt},\frac{i}{\sqrt{2}}(\alpha_1-\alpha_2)e^{iKt}).
\end{equation}

Considering the arguments of Bose broken symmetry, we assume that
two condensates are initially in the coherent state. So that the
wavefunction of the system at time $t$ can be written as
\begin{equation}
|\psi(t)>=e^{-N/2}\sum_{n,m=0}^{\infty}\frac{1}{\sqrt{n!m!}}(u_1e^{-iKt})^n(iu_2e^{iKt})^m\exp[-iE_{n,m}t]|n,m),
\end{equation}
where
\begin{equation}
u_1=\frac{1}{\sqrt{2}}(\alpha_1+\alpha_2),~~u_2=\frac{1}{\sqrt{2}}(\alpha_1-\alpha_2),
\end{equation}
\begin{equation}
N=|\alpha_1|^2+|\alpha_2|^2=|u_1|^2+|u_2|^2.
\end{equation}

\section{\label{sec:co}  Collapses and revivals of interference intensity
}

For convenience, we rewrite Hamiltonian (9) as follows
\begin{equation}
H=(\epsilon-3K)N_1+(\epsilon-3K)N_2+2KN_1^2+2KN_2^2-KN_1N_2.
\end{equation}
where $N_i=A_i^+A_i(i=1,2)$, $2KN_1^2$ and $2KN_2^2$ stand for
two-body hard-sphere collisions, $-KN_1N_2$ describes the
collision between the atoms of the two wells.

The dissipation is included by considering the master equation
\cite{kb,ji}
\begin{equation}
\frac{\partial\rho}{\partial
t}=-i[H,\rho]+\sum_{j=1,2}\gamma_j(2A_j\rho
A_j^+-A_j^+A_j\rho-\rho A_j^+A_j),
\end{equation}
where $\gamma_j~(j=1,2)$ denotes the dissipation or loss rate due
to some relaxation processes such as the coupling of the atoms in
the two wells with the environment. To solve Eq.(23), we can
introduce a transformation $\tilde{c}=\exp(iHt)c\exp(-iHt)$, then
Eq.(23) becomes
\begin{equation}
\frac{\partial\tilde{\rho}}{\partial
t}=\sum_{j=1,2}\gamma_j(2\tilde{A}_j\tilde{\rho}\tilde{A}_j^+
-\tilde{A}_j^+\tilde{A}_j\tilde{\rho}-\tilde{\rho}\tilde{A}_j^+\tilde{A}_j).
\end{equation}

The master equation (Eq.(24)) can be solved exactly for any chosen
initial state. In particular, when the atoms in the two wells are
initially in the coherent state (cs) $|\alpha_1,\alpha_2>$ or in a
Fock state (Fs) $|n,m>$, the corresponding density matrices are
given by, respectively
\begin{equation}
\tilde{\rho}^{(cs)}(t)=|\alpha_1e^{-\gamma_1t},\alpha_2e^{-\gamma_2t}>
<\alpha_1e^{-\gamma_1t},\alpha_2e^{-\gamma_2t}|,
\end{equation}
and
$$
\tilde{\rho}^{(Fs)}(t)=\sum_{l=0}^n\sum_{k=0}^m(e^{-2\gamma_1t})^{n-l}
(1-e^{-2\gamma_1t})^l(e^{-2\gamma_2t})^{m-k}(1-e^{-2\gamma_2t})^k$$
\begin{equation}
\times C_n^lC_m^k|n-l,m-k><n-l,m-k|,
\end{equation}
where
\begin{equation}
A_1|\alpha_1>=\alpha_1|\alpha_1>,~~A_2|\alpha_2>=\alpha_2|\alpha_2>,~~C_m^n=m!/n!(m-n)!.
\end{equation}

The Schr\"odinger picture field operator for the sum of the two
modes is $\psi=(A_1+A_2)/\sqrt{2}$, where the spatial dependence
has been suppressed \cite{gj,em}. The corresponding operator for
the intensity of the atomic pattern is $\psi^+\psi$ and its
time-varying expression can be obtained by the trace operator
$I(t)=Tr[\rho(t)\psi^+\psi]$. When the atoms in the two wells are
initially in the coherent state (cs) $|\alpha_1,\alpha_2>$, one
has
\begin{equation}
I^{(cs)}(t)=\frac{1}{2}\sum_{j=1,2}|\alpha_j|^2\exp(-2\gamma_jt)+|\alpha_1\alpha_2|
\exp[-\Gamma(t)]\cos\phi(t),
\end{equation}
where
\begin{equation}
\Gamma(t)=(\gamma_1+\gamma_2)t+2\sum_{j=1,2}|\alpha_j(t)|^2
\sin^2\frac{5}{2}Kt,
\end{equation}
\begin{equation}
\phi(t)=\beta+\sum_{j=1,2}(-1)^j|\alpha_j(t)|^2\sin 5Kt,
\end{equation}
\begin{equation}
|\alpha_j(t)|=|\alpha_j|\exp(-\gamma_jt),~~\alpha_1^*\alpha_2=|\alpha_1\alpha_2|\exp(-i\beta),
\end{equation}
where we have set $\alpha_1=|\alpha_1|\exp(i\phi_{\alpha_1})$ and
$\alpha_2=|\alpha_2|\exp(i\phi_{\alpha_2})$,
$\beta=\phi_{\alpha_1}-\phi_{\alpha_2}$.
 Eq.(28) can be
expanded as the form
$$I^{(cs)}(t)=\frac{1}{2}\sum_{j=1,2}|\alpha_j|^2\exp(-2\gamma_jt)
+|\alpha_1\alpha_2|\exp[-(\gamma_1+\gamma_2)t]$$
$$\times \sum_{m=-\infty}^{\infty}\sum_{n=-\infty}^{\infty}
\sum_{p=-\infty}^{\infty}\sum_{l=-\infty}^{\infty}I_m(|\alpha_1(t)|^2)
I_n(|\alpha_2(t)|^2)J_p(|\alpha_1(t)|^2)J_l(|\alpha_2(t)|^2)$$
\begin{equation}
\times \cos{\{\beta+[5m-5p+5n+5l]Kt}\},
\end{equation}
where $J_p(x)$ and $I_m(x)$ stand for the Bessel and modified
Bessel functions, respectively.

It is clear that when the dissipations are neglected
$(\gamma_j=0,j=1,2)$,  and we take the terms
\begin{equation}
m-p+n+l=0,
\end{equation}
we obtain a nonzero time-averaged value of the intensity of the
atomic pattern
$$I^{(cs)}=\frac{1}{2}(|\alpha_1|^2+|\alpha_2|^2)+|\alpha_1\alpha_2|\cos\beta$$
\begin{equation}
\times\sum_{m=-\infty}^{\infty}\sum_{n=-\infty}^{\infty}
\sum_{l=-\infty}^{\infty}
I_m(|\alpha_1|^2)I_n(|\alpha_2|^2)J_{m+n+l}(|\alpha_1|^2)J_l(|\alpha_2|^2).
\end{equation}
Eqs.(32) and (34) show that the intensity exhibits the revivals
and collapses. This phenomena also can be easily seen from Figure
1.

%-----------------------Fig 1------------------------------------------------
\begin{figure}[hbtp]
\centering
\includegraphics[width=0.5\textwidth,angle=0]{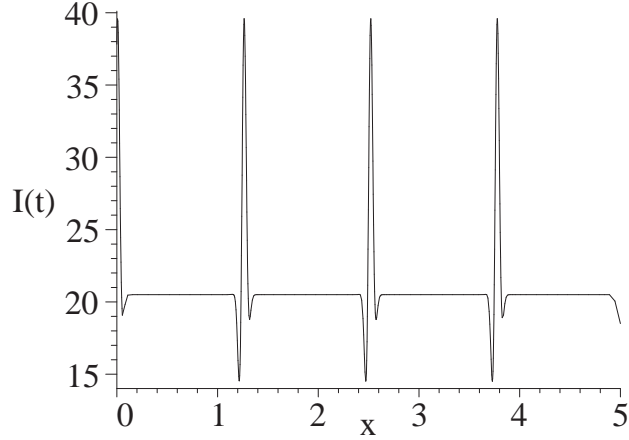}
\caption{Diagram of the time evolution of $I^{(cs)}(t)$. The time
is in unit of $x=Kt$. The result is shown for the case of
$\gamma_1=\gamma_2=0$, when the total number of atoms in the two
wells is $N=41$ with $|\alpha_1|=5$ and $|\alpha_2|=4$. Here,
$\beta=\pi/6$.}
\end{figure}
%------------------------------------------------------------------------------

On the other hand, when the atoms in the two wells are initially
in a Fock state (Fs) $|n,m>$, one has
\begin{equation}
I^{(Fs)}(t)=\frac{1}{2}\sum_{l=0}^n\sum_{k=0}^m(n+m-k-l)(e^{-2\gamma_1t})^{n-l}
(1-e^{-2\gamma_1t})^l(e^{-2\gamma_2t})^{m-k}(1-e^{-2\gamma_2t})^kC_n^lC_m^k,
\end{equation}

%-----------------------Fig 2------------------------------------------------
\begin{figure}[hbtp]
\centering
\includegraphics[width=0.5\textwidth,angle=0]{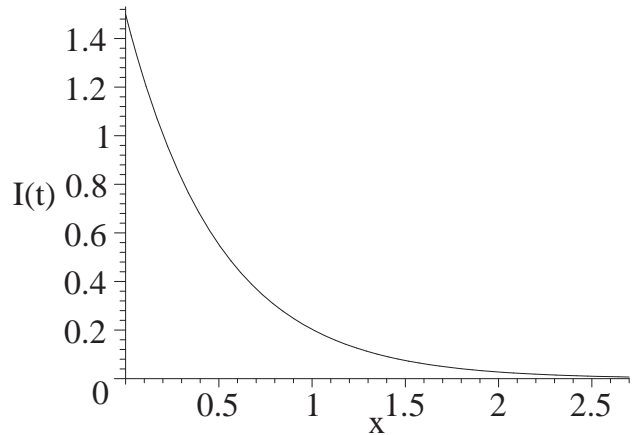}
\caption{Diagram of the time evolution of $I^{(Fs)}(t)$. The time
is in unit of $x=\gamma_1 t=\gamma_2 t$, we have set
$\gamma_1=\gamma_2$. }
\end{figure}
%------------------------------------------------------------------------------
which shows that the intensity does not exhibit collapses and
revivals (see Fig. 2).

\section{\label{sec:fi}  Fidelity of Interference Between the atoms in the two
wells}

The fidelity and its loss rate of interference between the atoms
in the two wells may be characterized by\cite{sm}
\begin{equation}
\tilde{F}=<\psi_0|\tilde{\rho}(t)|\psi_0>,
\end{equation}
\begin{equation}
\tilde{L}=-<\psi_0|\frac{\partial\tilde{\rho}}{\partial
t}|\psi_0>|_{t=0},
\end{equation}
where $|\psi_0>$ is the initial state of the system,
$\tilde{\rho}(t)$ and $\partial\tilde{\rho}/\partial t$ satisfy
Eq.(24).

We now turn to study the fidelity of interference between the
atoms in the two wells. When the atoms in the two wells are
initially in the coherent state $|\alpha_1,\alpha_2>$, the
corresponding fidelity of interference is given by
\begin{equation}
\tilde{F}^{(cs)}=\exp[-|\alpha_1|^2(1-e^{-\gamma_1t})^2
-|\alpha_2|^2(1-e^{-\gamma_2t})^2].
\end{equation}
Similarly, for the initial Fock state, one has
\begin{equation}
\tilde{F}^{(Fs)}=\exp[-2n\gamma_1t-2m\gamma_2t].
\end{equation}
Diagrams of the time evolution of $\tilde{F}^{(cs)}(t)$ and
$\tilde{F}^{(Fs)}(t)$ see Fig. 3.
%-----------------------Fig 3------------------------------------------------
\begin{figure}[hbtp]
\centering
\includegraphics[width=0.5\textwidth,angle=0]{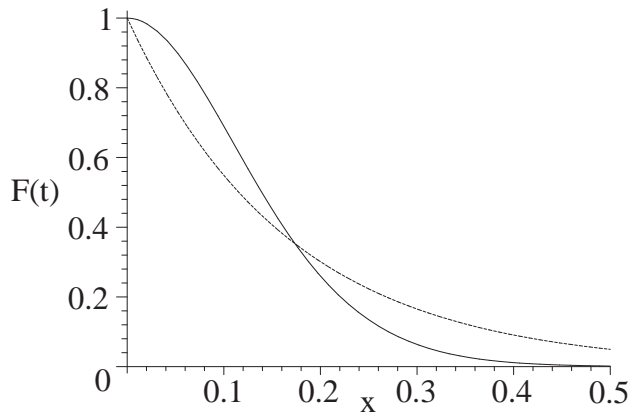}
\caption{Diagrams of the time evolution of
$\tilde{F}^{(cs)}(t)$(solid line) and $\tilde{F}^{(Fs)}(t)$(dash
line). The time is in unit of $x=\gamma_1 t=\gamma_2 t$ .For
simplicity, we have set $\gamma_1=\gamma_2$. The total number of
atoms in the two wells is $N=41$ with $|\alpha_1|=5$ and
$|\alpha_2|=4$. The Fock state is supposed in $|n,m>=|1,2>$.}
\end{figure}
%------------------------------------------------------------------------------

Because the large particle number of the two condensates implies
small $\gamma_j\tau_{Fid}$, we can set
$1-\exp[-\gamma_j\tau_{Fid}]\cong\gamma_j\tau_{Fid}$, such that
\begin{equation}
\tilde{F}^{(cs)}\cong \exp[-(|\alpha_1|^2\gamma_1^2
+|\alpha_2|^2\gamma_2^2)t^2],
\end{equation}
for short time. The resulting fidelity times are then
\begin{equation}
\tau_{Fid}^{(cs)}\cong[|\alpha_1|^2\gamma_1^2+|\alpha_2|^2\gamma_2^2]^{-1/2},~~
\tau_{Fid}^{(Fs)}=(2n\gamma_1+2m\gamma_2)^{-1},
\end{equation}
which show that the fidelity time is not only related to the
initial state of the system, but also to the dissipation
parameters.

Furthermore, we can get the fidelity loss rates:
\begin{equation}
\tilde{L}^{(cs)}=0,~~\tilde{L}^{(Fs)}=2n\gamma_1+2m\gamma_2,
\end{equation}
which indicate that when the atoms in the two wells are initially
in the coherent state, the fidelity loss rate of interference is
zero, but for the initial Fock state, $\tilde{L}^{(Fs)}$ is
related to the initial particle number of the system and the
dissipation parameters, but not to the collision parameters.

\section{\label{sec:con} Conclusions}

We have studied interference between the halves of a double-well
trap containing a BEC. It is found that when the atoms in the two
wells are initially in the coherent state, the intensity exhibits
collapses and revivals, but it does not for the initial Fock
states. The interference intensity is affected by the collision
and dissipation, but for the initial Fock state, it is only
related to the dissipation. Whether the initial states are in the
coherent states or in a Fock states, the fidelity time has nothing
to do with collision. For the initial coherent states, the
fidelity loss rate is zero, but for the initial Fock states, it is
determined by the initial particle number of the system and
dissipation. This shows that interference and its fidelity can be
adjusted experimentally by properly preparing the number and
initial states of the system.

It is pointed out that the recent realization of a
superfluid-Mott-insulator phase transition in a gas of ultra-cold
atoms in an optical lattice\cite{vv} is very similar to the state
preparation assumed in this paper, we hope our results obtained
above will be useful to study Mott insulator phase transition in
the future.

\end{document}